\documentclass[amsmath,amssymb,aip,jcp,10pt,twocolumn]{revtex4-1}

\usepackage{dcolumn}
\usepackage{graphicx}
\usepackage{graphics}
\usepackage[usenames]{color}
\usepackage{soul}
\usepackage{amsmath,amssymb}
\usepackage[mathscr]{euscript}

\begin{document}

\author{Jason N. Byrd}
\email{byrd.jason@ensco.com}
\affiliation{Quantum Theory Project, University of Florida, Gainesville, Florida 32611, USA}
\affiliation{ENSCO, Inc., 4849 North Wickham Road, Melbourne, Florida 32940, USA}
\author{Jesse J. Lutz}
\email{jesse.lutz.ctr@afit.edu}
\thanks{This research was supported in part by an appointment to the Faculty Research Participation Program at 
U.S. Air Force Institute of Technology (AFIT),
administered by the Oak Ridge Institute for Science and Education through an interagency agreement between the
U.S. Department of Energy and AFIT.}
\affiliation{Quantum Theory Project, University of Florida, Gainesville, Florida 32611, USA}
\author{Yifan Jin}
\affiliation{Quantum Theory Project, University of Florida, Gainesville, Florida 32611, USA}
\author{Duminda S. Ranasinghe}
\affiliation{Quantum Theory Project, University of Florida, Gainesville, Florida 32611, USA}
\author{John A. Montgomery, Jr.}
\affiliation{Department of Physics, University of Connecticut, Storrs, Connecticut 06269, USA}
\author{Ajith Perera}
\affiliation{Quantum Theory Project, University of Florida, Gainesville, Florida 32611, USA}
\author{Xiaofeng F. Duan}
\affiliation{Air Force Research Laboratory DoD Supercomputing Resource Center, Wright-Patterson Air Force Base, OH 45433, USA}
\affiliation{Air Force Institute of Technology, Wright-Patterson Air Force Base, Ohio 45433, USA}
\author{Larry W. Burggraf}
\affiliation{Air Force Institute of Technology, Wright-Patterson Air Force Base, Ohio 45433, USA}
\author{Beverly A. Sanders}
\affiliation{Department of Computer and Information Science and Engineering, University of Florida, Gainesville, Florida 32611, USA}
\affiliation{Quantum Theory Project, University of Florida, Gainesville, Florida 32611, USA}
\author{Rodney J. Bartlett}
\email{rodbartl@ufl.edu}
\affiliation{Quantum Theory Project, University of Florida, Gainesville, Florida 32611, USA}

\title{Predictive coupled-cluster isomer orderings for some Si${}_n$C${}_m$ ($m, n\le 12$) clusters; 
A pragmatic comparison between DFT and complete basis limit coupled-cluster benchmarks.}

\begin{abstract}
The accurate determination of the preferred ${\rm Si}_{12}{\rm C}_{12}$ isomer
is important to guide experimental efforts directed towards synthesizing SiC
nano-wires and related polymer structures which are anticipated to be highly
efficient exciton materials for opto-electronic devices.  In order to definitively
identify preferred isomeric structures for silicon carbon nano-clusters, highly
accurate geometries, energies and harmonic zero point energies have been
computed using coupled-cluster theory with systematic extrapolation to the
complete basis limit for set of silicon carbon clusters ranging in size from
SiC$_3$ to ${\rm Si}_{12}{\rm C}_{12}$.  It is found that post-MBPT(2)
correlation energy plays a significant role in obtaining converged relative
isomer energies, suggesting that predictions using low rung density functional
methods will not have adequate accuracy.  Utilizing the best composite
coupled-cluster energy that is still computationally feasible, entailing a 3-4
SCF and CCSD extrapolation with triple-$\zeta$ (T) correlation, the {\it closo}
${\rm Si}_{12}{\rm C}_{12}$ isomer is identified to be the preferred isomer in
support of previous calculations [J. Chem. Phys. 2015, 142, 034303].
Additionally we have investigated more pragmatic approaches to obtaining
accurate silicon carbide isomer energies, including the use of frozen natural
orbital coupled-cluster theory and several rungs of standard and double-hybrid
density functional theory.  Frozen natural orbitals as a way to compute post
MBPT(2) correlation energy is found to be an excellent balance between
efficiency and accuracy.  
\end{abstract}

\maketitle

\section{Introduction}
Recent advancements in the nanoscale design, precision measurement, and coherent
control of silicon-based materials are driving rapid developments in
electronics,\cite{ZDMrmp2013} photonics\cite{LKFnp2010},
spintronics,\cite{Jnm2013,MLRnm2015} and excitonics.\cite{SRnm2006,KKKnm2006}
Among the most promising materials, in terms of manufacturing cost,
physical durability, and engineering flexibility, is silicon carbide (SiC).  SiC
has a wide band gap, high thermal conductivity, high breakdown electric field,
high saturated electron drift velocity, and it is radiation resistant.  This makes
it an excellent refractory material for devices which must endure extreme conditions, 
such as those present in nuclear reactors or interstellar space.  Hence, if it can be
created efficiently silicon carbide materials are likely to supplant the bulk silicon crystals
currently used in manufacturing microelectronic, photovoltaic, and microelectromechanical system technologies.  
Furthermore, the diversity of properties exhibited by its many
accessible polytypes and nanostructures provides a generous design flexibility
which may enable development of novel materials suitable for mainstream
production of excitonics, spintronics, and photonics
devices.\cite{RBBoe2013,MLRnm2015}

Large-scale SiC materials design is still at an early stage from the standpoint
of both the experimentalist and the theoretician.  Preparation of
one-dimensional Si-C binary core/shell nanowires has for some time been
reported,\cite{ZZPcpl2000,RPSjcg2004,YATam2005,AVCapl2012,OLRjcg2013,NNRjl2015,GRmcp2014,Rprb2005}
but only very recently was a procedure developed for obtaining
highly-crystalline 2-D SiC sheets with nanometer thickness.\cite{CCXn2016} The
latter achievement was inspired by earlier theoretical predictions of the
stability of single-layer SiC,\cite{MMTnm2007,BTCprb2010,LZZn2014} and this
exemplifies the synergy between theory and experiment that has been a common
thread in many modern nanoscale materials science advancements.
Such breakthroughs stimulate further development of efficient methodologies for
the bottom-up theoretical design and engineering of novel nanoscale SiC
materials.  At the other end of the size spectrum, SiC containing nano-clusters 
and small molecules are prevalent in the interstellar medium.  Because of the 
difficulties in experimental replication of the interstellar environment, 
there is a great demand for theory to provide any comparable molecular data.

Nearly all stable Si$_n$C$_m$ molecules do not resemble their bulk SiC counterpart.
Solid-state SiC consists of mostly extended chains of alternating Si and C atoms, 
while Si$_n$C$_m$ clusters tend to be segregated into carbon- and silicon-rich
regions, with sporadic linkage through Si-C bonds. When $n$ is small, silicon 
link to the periphery as single atoms or clusters, and when $n$ is large, the
silicon network spans the carbon moiety. The relative stability of Si$_m$C$_n$
molecules tends to maximize with $n=m$, and, in particular, the
Si$_{12}$C$_{12}$ molecule has become motivating due to its ability to form
different polymer chains as well as 2D and 3D networks by Si-Si
bonding.\cite{DBjcp2016} However, predicting in advance the most
energetically-favorable arrangement of the atoms in such systems can be very
challenging.

{\it Ab initio} methods are an excellent tool for sampling the configuration
space of a molecular cluster to find the global minimum.  As the most
accurate methods are extremely computationally taxing, identification of more efficient
approaches is desired and this can be done by comparison with benchmark values, if possible.
Unfortunately the structural details of large Si$_n$C$_m$ clusters are not yet
documented, but parameters for smaller clusters have for some time been known from
interstellar and laboratory spectroscopy.  These measurements form the basis for
a systematic study which may help determine the most efficient methods for
studying larger clusters or perhaps even periodic SiC systems. 

The first observation of visible-range interstellar absorption bands
corresponding to Si$_n$C$_m$ clusters were reported in 1926 by
Merrill\cite{Merrill} and Sanford,\cite{Sanford} and these bands were later
correctly attributed to the SiC$_2$ molecule by Kleman.\cite{Kleman} Other small
silicon carbide clusters have been observed using radio astronomy including
SiC,\cite{Cernicharo} rhomboidal SiC$_3$,\cite{Apponi} and linear
SiC$_4$.\cite{Ohishi} Gas-phase IR spectra measured in the laboratory helped
further characterize ground-state Si$_n$C$_m$ geometries, providing definitive
structural parameters for SiC,\cite{BernathPRL1988} triangular
SiC$_2$,\cite{SGjcp1985,Michalopoulous,PGjcp1991} triangular
SiC$_2$,\cite{PGSjcp1990} linear SiC$_4$,\cite{WGjcp1992,VGPcpl1995} SiC$_n$ ($n
> 4$),\cite{DWRGjcp1999,DWRGjcpa2000,LRGjcp2014a} triangular
Si$_2$C,\cite{KHFMjpc1983,RBBjcp2015} rhombic Si$_2$C$_2$,\cite{PGRGjcp1995}
linear SiC$_3$Si,\cite{PGjcp1994,VGPjcp1994,Thorwirth} linear
SiC$_4$Si,\cite{PRGjcp1997} Si$_2$C$_5$,\cite{LRGjcp2014b} rhomboidal
Si$_3$C,\cite{PGjcp1992} pentagonal Si$_3$C$_2$,\cite{PGjcp1996} and
silicon-rich Si$_n$C (with $n$ = 3--8) \cite{TruongPCCP2015} and Si$_n$C$_m$
(with $n + m = 6$) clusters.\cite{Savoca} Computational modeling has also been
performed by several groups in order to describe small ground-state Si$_n$C$_m$
clusters,\cite{Rjcp1992,Rjcp1994,GSjcp1984,GSjcp1985,GScpl1985,AGSjcp1990,
JXWjms2002,JXWjsc2003,JXWjpca2003,JXWjms2003a,JXWjms2003b,HSjcp2008,SYHHepjd2010,
ZLZjpcm2011,ETpe2000,LLZctc2011,HJjcp1996,ZFMjcp1996,AJIpe2016}
heterofullerenes,\cite{momeni2010,huda2008,scipioni2011,yu2012} cage
structures,\cite{Pochet2010} silafullerenes,\cite{tillmann2015} and
graphene-silicene bilayers.\cite{NASapl2013,BNAprb2014}

In 2010, several low-lying isomers of the Si$_n$C$_m$ ($m,n \le 4$) clusters
were studied to determine the most energetically stable ground-state
structures.\cite{DWBcms2010} At that time the reported structures reinforced all
known spectroscopic interpretations, but since then Truong et al. measured a
ground-state geometry for Si$_4$C in disagreement with the
predictions.\cite{TruongPCCP2015}  This discrepancy also called into
question the reliability of other calculations performed on medium-sized
Si$_n$C$_m$ clusters ($m,n \le 12$).\cite{DBHm2013} Since those results led to
the intriguing prediction of the existence of a stable, cage-like {\em
closo}-Si$_{12}$C$_{12}$ structure,\cite{DBjcp2015} it is of great interest to
know whether the reported DFT isomer energy-ordering is correct and {\em
closo}-Si$_{12}$C$_{12}$ is in fact the most thermodynamically stable isomer.

This work has two related motivations.  Firstly, before embarking on an effort
to synthesize the {\em closo}-Si$_{12}$C$_{12}$ molecule and its polymeric
extensions,\cite{DBjcp2016} one needs to definitively confirm that it is, in
fact, the most stable isomer.  If this initial prediction is proven to be in
question, more difficult and less efficient kinetic synthesis approaches will
become necessary.  Secondly, once highly-accurate coupled-cluster results are
obtained, the effectiveness of more pragmatic approaches for the treatment of
large Si$_n$C$_m$ clusters can be explored.  In this way it is desirable to
identify a DFT-based method which can offer a much more efficient yet reliable
alternative for determining the lowest-energy isomer searches and other
computational predictive studies on Si$_n$C$_m$ clusters and their polymeric
analogue.

\section{Electronic Structure Calculations}

We performed all reported electronic structure calculations 
using the serial ACESII,\cite{acesii} and parallel ACESIII,\cite{acesiii2008}
Aces4,\cite{aces4-electronic} GAMESS,\cite{gamess1993,gamess2005}
ORCA,\cite{orca} and NWCHEM\cite{nwchem2010} {\it ab initio} quantum chemistry
packages.  Correlation calculations were performed using the ACES family of {\it
ab initio} programs on the NAVY DSRC Cray XC30 Shepard and ARL DSRC Cray XC40
Excalibur.  Density functional theory (DFT) calculations were performed on the
University of Florida HiPerGator cluster and NAVY DSRC Cray XC30 Shepard using
GAMESS and NWCHEM, while double-hybrid density functional theory calculations
were performed using ORCA on a local work station.  
Correlation and DFT calculations performed here use Dunning's correlation
consistent family of basis sets (cc-pV$n$Z and tight function variant
cc-pV($n$+d)Z, $n$=D,T,Q,5) optionally with diffuse functions ({aug-}) and with
weighted core-valence functions
(cc-pwCV$n$Z).\cite{dunning1989,woon1993,dunning2001}  All double-hybrid (DH)
DFT calculations used the resolution-of-the-identity (RI) approximation in the
DFT\cite{weigend2009} and MBPT(2) calculation using the
def2-QZVP\cite{weigend2003} basis.  Correlation calculations in this work assume
the frozen-core approximation where all carbon $1s$ and silicon $1s2s2p$
orbitals are frozen at the SCF level and dropped from the correlation space
unless explicitly stated otherwise.  Throughout this work, a very tight grid is
employed, the \texttt{JANS=2} grid in GAMESS and \texttt{xfine} in NWCHEM.
Geometry optimizations and single point energy calculations were performed using
a large variety of methods including second-order many-body perturbation theory
with the M{\o}ller-Plesset partitioning (MBPT(2)), linear coupled-cluster
doubles (LCCD),\cite{bartlett1981,taube2009,bartlett2010,byrd2015-c}
coupled-cluster theory with singles and doubles (CCSD),\cite{purvis1982} and
CCSD with perturbative triples\cite{urban1985,raghavachari1989,watts1993}
(CCSD(T)), and some DFT functionals.  Throughout MAX refers to the maximum
unsigned error, MUEs as the mean unsigned error, and RMS as the root mean square
of the error.

When computing the relative isomer energies for large molecular systems using
conventional {\it ab initio} methods, it is convenient to take advantage of
energy additivity to analyze the calculated energies by orders of perturbation
theory.  This enables the systematic inclusion of higher-order estimates of the
correlation energy, often illustrated by the {\it ab initio} hierarchy,
MBPT(2)$<$LCCD$<$CCSD$<$CCSD(T)$<\dots<$FCI. As a function of basis set the
analogous hierarchy is D$\zeta <$T$\zeta <$Q$\zeta <\dots<\infty\zeta$, where
the latter term is the infinite basis set idealization.  To obtain the best
possible isomer energies for our available computational resources, the
individual energy contributions computed with a given basis set can be separated
and added to form an aggregate total energy.  Summarizing our notation, the
MBPT(2) correlation energy is represented by 
\begin{equation}
\Delta {\rm MBPT(2)}=E({\rm MBPT(2)})-E({\rm SCF})
\end{equation}
with higher order contributions similarly defined as 
\begin{equation}
\Delta {\rm CCSD}=E({\rm CCSD})-E({\rm SCF}),
\end{equation} 
\begin{equation}
\Delta {\rm (T)}=E({\rm CCSD(T)})-E({\rm CCSD}),
\end{equation} 
and 
\begin{equation}
\Delta {\rm CC} = E({\rm CCSD(T)})-E({\rm MBPT(2)})
\end{equation}
as the post MBPT(2) correlation energy.  

As an alternative to including the correlation energy from a small basis or
using a large amount of computational resources to obtain  large basis results,
it is possible to approximate the largest basis set correlation energies at the
cost of a small basis using the frozen natural orbital (FNO)
method.\cite{taube2005,taube2008}  The FNO scheme is a way to truncate the
virtual orbital space in a post-MBPT(2) (CCSD and beyond) calculation with a
$p^4$ ($p$ is the amount of virtual space truncated) computational savings.  As
is evident by the polynomial dependence on the number of virtual orbitals,
truncating even small portions of the virtual space can lead to tremendous
computational savings.  Our goal of calculating accurate isomerization energies
is an ideal target for the FNO method\cite{byrd2015-a} due to the favorable
cancellation of error to be obtained from the energy differences of structurally
similar isomers.

The FNO method works by ordering the virtual orbitals based on the MBPT(2)
virtual-virtual density occupation numbers then truncating the remaining space
based on an occupation number threshold criteria; details of this ordering and
truncation can be found in the FNO
references.\cite{taube2005,taube2008,deprince2012}  In our
experience\cite{taube2005,taube2008,byrd2015-a} virtual space truncation using a
$1\times 10^{-4}$ threshold (which usually means a removal of $~30-35\%$ of the
virtual orbitals) is a good balance between accuracy and
efficiency.\footnote{Care must be made to not select a threshold that would
incorrectly remove degenerate virtual orbitals.}  The general energy
decomposition notation for FNO energies is 
\begin{equation}
\Delta_{\rm FNO} {\rm (T)}=E_{\rm FNO}({\rm CCSD(T)})-E_{\rm FNO}({\rm CCSD}),
\end{equation}
and
\begin{equation}
\Delta_{\rm FNO} {\rm CC} = E_{\rm FNO}({\rm CCSD(T)})-E({\rm MBPT(2)}).
\end{equation}
As a note, the $\Delta_{\rm FNO} {\rm (T)}$ energy is computed using the
truncated virtual FNO $T$ amplitudes, which incurs only a small
error.\cite{deprince2012,byrd2015-a}

It is well known that the basis set convergence of post-MBPT(2) correlation is
much faster than the SCF and second-order contribution.\cite{platts2013} Because
each contribution to the total energy has a different basis set dependence, we
can estimate the effects of taking the total energy to the complete basis set
limit (CBS) by examining the basis set dependence of each contribution ($E({\rm
SCF})$, $\Delta {\rm MBPT(2)}$, $\Delta {\rm CCSD}$, and $\Delta {\rm (T)}$)
individually.  Extrapolation of smaller basis set energies is a very effective
way to estimate the CBS limit without requiring the use of very large basis sets
to directly obtain an energy near the CBS limit.  We employ the linear
extrapolation formula of Schwenke\cite{schwenke2005}
\begin{align}\label{scfcbseqn}
E^n_\infty({\rm SCF}) &=E_{n-1}({\rm SCF})+F^{\rm SCF}_{n-1,n}(E_n({\rm SCF})-E_{n-1}({\rm SCF})) \\
\label{cccbseqn}
\Delta^n_\infty{\rm M} &=\Delta_{n-1}{\rm M}+F^{\rm M}_{n-1,n}(\Delta_n{\rm M}-\Delta_{n-1}{\rm M}) 
\end{align}
and the  cubic extrapolation formula of Helgaker {\it et al.}\cite{helgaker1997}
\begin{equation}\label{cbs}
\Delta^n_\infty {\rm MBPT(2)}=
\frac{n^3 \Delta_n {\rm MBPT(2)} - (n-1)^3 \Delta_{n-1} {\rm MBPT(2)}}{n^3-(n-1)^3}
\end{equation}
to obtain CBS estimates of the SCF, MBPT(2) and higher order ($\Delta {\rm
CCSD}$, and $\Delta {\rm (T)}$) contributions to the total energy.  Here
$E_n({\rm M})$, $\Delta_n {\rm MBPT(2)}$ and $\Delta_n({\rm M})$ refer to that
energy contribution computed with the cc-pV$n$Z basis while $F^{\rm M}_{n-1,n}$
is a tabulated quantity.\cite{schwenke2005}   There are a number of ways to
assemble the CBS estimates of the SCF and correlation energies which vary in
accuracy and computational cost.  Based on what is currently computationally
feasible, we define a best estimate of the CBS extrapolated energy to be
\begin{equation}\label{eqcomp}
 E_{\rm CBS}({\rm Best}) = 
 E^5_\infty ({\rm SCF}) + \Delta^4_\infty ({\rm CCSD}) + \Delta_3 ({\rm T}) + {\rm ZPE}
\end{equation}
where ${\rm ZPE}$ is the harmonic zero point energy (ZPE) computed at the
MBPT(2)/cc-pVTZ level of theory.  For smaller clusters we additionally
investigate the $\Delta {\rm CCSD}$ core-valence correlation energy ($\Delta{\rm
CV}$ using the corresponding cc-pwCV$n$Z basis set) and compute an MBPT(2)-level
fine-structure relativistic correction ($\Delta{\rm FS}$).  Other, more
approximate estimates for the CBS total energy employed in this work are
\begin{widetext}
\begin{align}
\label{mbptcbs}
{\rm MBPT(2)}::\Delta_2 {\rm CC} &= E^4_\infty ({\rm SCF}) + \Delta^4_\infty ({\rm MBPT}) + \Delta_2 {\rm CC} + {\rm ZPE}, \\
\label{fnocbs}
{\rm MBPT(2)}::\Delta_{\rm FNO}{\rm CC} &= E^4_\infty ({\rm SCF}) + \Delta^4_\infty ({\rm MBPT}) + \Delta_{\rm FNO} {\rm CC} + {\rm ZPE}, \\
\label{cccbs2}
{\rm CCSD}::\Delta_n ({\rm T}) &= E^4_\infty ({\rm SCF}) + \Delta^4_\infty ({\rm CCSD}) + \Delta_n {\rm (T)} + {\rm ZPE}, \\
\label{cccbs3}
{\rm CCSD}::\Delta_{\rm FNO}({\rm T}) &= E^4_\infty ({\rm SCF}) + \Delta^4_\infty ({\rm CCSD}) + \Delta_{\rm FNO} {\rm (T)} + {\rm ZPE}.
\end{align}
Each of which uses a cc-pV\{T,Q\}Z extrapolation to estimate the SCF and leading
order correlation energy CBS limit, with higher order correlation contributions
included at either the cc-pVDZ, FNO/cc-pVTZ or full cc-pVTZ level.
\end{widetext}

A computationally attractive method for computing the energies of large
molecules is DFT, which has had a long history of competition with wave-function
theory.  An extensive examination of the performance of density functional
theory in the calculation of structures and energies of Si$_n$C$_m$ clusters is
beyond the scope of this work.  However we have included a survey of relative
isomer energies computed by a number of common functionals including
B3LYP\cite{becke1993,stephens1994} (using VWN5), cam-QTP(0,0),\cite{verma2014}
cam-QTP(0,1),\cite{jin2016} $\omega$B97X-D2,\cite{chai2004,chai2008}
M06-2X,\cite{zhao2006}, and M11\cite{peverati2011}  in our results to provide a
point of comparison for future efforts.  We also
provide values computed using the B2-PLYP,\cite{grimme2006a,schwabe2008}
B2GP-PLYP,\cite{karton2008} DSD-BLYP,\cite{kozuch2010}
DSD-PBEP86,\cite{kozuch2011} and PWPB95\cite{goerigk2011} DH-DFT.
The -D2 (2006) and -D3 (2010) Grimme dispersion
corrections\cite{grimme2006b,grimme2010} were employed as denoted above.

\section{Computational Results and Discussion}

In this study we develop and apply accurate yet efficient computational
protocols for the determination of structural parameters and energetic rankings
of low-lying Si$_n$C$_m$ isomers.  To this end, we set forth several objectives:
(1) establish a list of approaches which efficiently determine structural
parameters for small Si$_n$C$_m$ clusters, (2) verify the level of accuracy
achieved by the benchmark composite method (Eq. \ref{eqcomp}) when used to
compute isomer energy differences for small Si$_n$C$_m$ clusters, (3) identify
those computational approaches that provide a qualitatively correct
energy-ordering of optimized structures of small and medium Si$_n$C$_m$ isomers
by comparing with benchmark-quality results, and (4) recommend pragmatic
procedures for the determination of qualitatively correct isomer energy ordering
which scales from small through medium Si$_n$C$_m$ clusters, up to and including
the target Si$_{12}$C$_{12}$ system.

\subsection{Efficiently Optimizing Geometries of Si$_n$C$_m$ Clusters with $n, m\le 2$}
%\subsection{Efficiently Optimizing Geometries of Small Si$_n$C$_m$ Clusters}

Initially we investigated the accuracy of various combinations of method and
basis set for describing the ground-state structural parameters of the four
smallest ($n, m\le 2$) Si$_n$C$_m$ clusters.  The results of geometry
optimizations performed on the $^3$X surface of SiC and on the $^1$X surfaces of
SiC$_2$, Si$_2$C, and Si$_2$C$_2$ are compared in Table \ref{tabsmalls} where
the MUEs of the collective optimized bond lengths and collective optimized bond
angles are reported for ground-state SiC, SiC$_2$, Si$_2$C, and Si$_2$C$_2$
clusters.  Benchmark bond distances and bond angles were taken to be those
resulting from optimization at the highest level of theory considered,
CCSD(T)/aug-cc-pV(Q+d)Z.  The full set of MUEs determined 
for each system at the various levels of theory can be found in the 
supplemental material.\cite{jcpsupp}

When basis sets of D$\zeta$-quality are employed, every wave-function method
considered produced large MUEs of over 3.0 pm and 7.0 degrees for the bond
lengths and angles, respectively.  At this basis set level the DFT functionals
outperform the wave-function methods we tested by at least a factor of two,
producing MUEs of 0.7--1.5 pm and 0.3--1.1 degrees.  The computational
efficiency of DFT makes small-basis B3LYP, M11, and $\omega$B97X-D good
candidates for obtaining structural parameters when large-basis wave-function
approaches are unaffordable.  It is yet to be seen whether these methods also
produce accurate isomer energy orderings.

Improving the basis sets to T$\zeta$ quality caused uniform improvement in the
performance of all relevant wave-function methods, with associated MUEs
decreasing to within 1.0 pm and 0.7 degrees for bond lengths and angles,
respectively.  In contrast, the MUEs of the DFT methods did not uniformly
improve when increasing the basis set size.  This introduces uncertainty
regarding the expected performance of DFT methods for larger systems, since the
number of basis functions necessarily grows relative to the system size.
However, this small basis accuracy illustrates that for such DFT methods the
basis set is readily saturated.  Of the more reliable wave-function methods
tested, MBPT(2) is the most computationally tractable, identifying it as a
leading candidate for use in optimizations.

In general, increasing the basis set size beyond the standard cc-pVTZ produced
diminishing returns.  The inclusion of diffuse or tight-$d$ functions is shown
in Table \ref{tabsmalls} to be non-essential at both the D$\zeta$ or T$\zeta$
level for all methods tested, resulting in improvements in MUEs of, at best,
0.1--0.3 pm or 0.1--0.3 degrees for the bond lengths and angles, respectively.
Similarly, improving the basis set from T$\zeta$- to the more expensive
Q$\zeta$-quality generally did not produce a worthwhile reduction in the MUEs.
In light of these findings, the optimization approaches employed in the
remainder of this work are limited to having either cc-pVDZ or cc-pVTZ basis
sets. The MBPT(2)/cc-pVTZ optimized geometries for all Si$_{m}$C$_{n}$ 
clusters with $m \leq n \leq 4$ can be found in the supplemental material.\cite{jcpsupp}

\begin{figure}
\includegraphics[width=0.4\paperwidth]{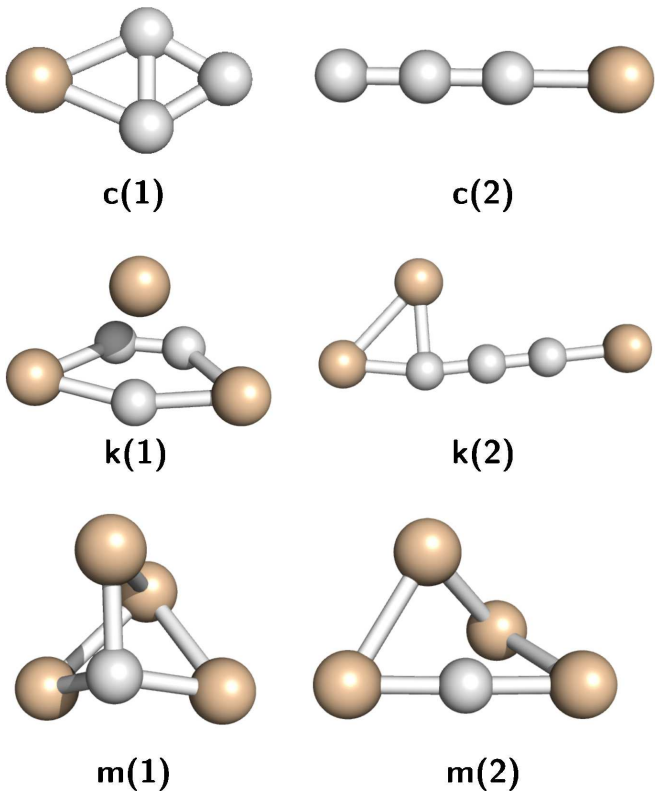}
\caption{The lowest-lying isomers of SiC$_3$, Si$_3$C$_3$, and Si$_4$C, as
 optimized using MBPT(2)/VTZ. 
\label{figmeds}
}
\end{figure}

\subsection{Validating a Composite Method for Generating Benchmark Values}

Small Si$_n$C$_m$ clusters were also used to validate the accuracy of the
composite method outlined in Eq. \ref{eqcomp}, which is a useful tool for
benchmarking the performance of other approaches.  Our strategy for this task is
to examine individual contributions to the total isomer energy differences and
study their convergence behavior as a function of basis set size.  The SiC$_3$,
Si$_3$C$_3$, and Si$_4$C systems are considered in this way, since these three
systems have been previously identified in the literature as having contentious
lowest-energy isomer configurations.\cite{DWBcms2010,SteglichAJ2015}  In
conformance with the notation of Duan {\it et al.},\cite{DWBcms2010} the two
lowest-lying isomeric structures are labeled as c1 and c2 for SiC$_3$, k1 and k2
for Si$_3$C$_3$, and m1 and m2 for Si$_4$C. The corresponding geometries were
optimized at the MBPT(2)/cc-pVTZ level and are sketched in Figure \ref{figmeds}
for reference. 

Leading contributions to the isomer energy differences, as obtained for the
SiC$_3$, Si$_3$C$_3$, and Si$_4$C isomer pairs, are presented in Table
\ref{tabmeds}.  Focusing first on SiC$_3$, only small changes are found when
moving from the cc-pVDZ to the cc-pVTZ basis sets for most of the individual
energies, with the exception of the SCF and $\Delta {\rm CCSD}$ quantities.  By
then stepping back to examine the $\Delta {\rm CCSD}$ convergence behavior for
all three molecular systems, it is clear that Q$\zeta$-quality calculations are
generally required before values are converged to within 1 kcal/mol of the CBS
limit.  Due to the small extra computational effort involved, \{TZ,QZ\}
correlation energy extrapolations were routinely performed, as were large-basis
SCF calculations.

Considering now the magnitude of the various energy differences computed at the
cc-pVDZ level, several appear relatively small.  For this set of systems, the
$\Delta{\rm CV}$ and $\Delta{\rm FS}$ contributions never contribute more than
0.5 kcal/mol to their respective energy differences.  Since the relative size of
both of these corrections is expected to decrease in moving to larger SiC
clusters, it is probably safe to omit them.  Finally, the ZPE values are
reasonably converged at the MBPT(2)/cc-pVDZ level, but can also be conveniently
be obtained at the MBPT(2)cc-pVTZ level following a geometry optimization at the
same level.  On the basis of these observations for small clusters, it appears
that the composite method given by Eq. \ref{eqcomp} will provide isomer enthalpy
differences approaching $\sim 1$ kcal/mol for larger SiC clusters. 

\subsection{Accurately ordering the relative isomer energy for Si$_n$C$_m$ clusters with $n, m\le 2$}
%\subsection{Accurately Ordering Small Si$_n$C$_m$ Isomers by Energy}

Another important consideration when developing an approach for searching for
the lowest-energy isomer is whether the level of theory used for geometry
optimizations will also provide an appropriate energy-ordering of the optimized
structures. Often when only the structure of the global minimum is of
interest, all other stationary points are discarded once the leading candidate
is identified. This can be problematic however, since the energy-ordering
resulting from the optimization method can be misleading, even in the case where
very accurate structural parameters are returned.  To test this we focus on the
challenging case of the two lowest-lying structures of Si$_4$C, which are known
to interchange depending on the level of theory employed.

In Table \ref{tabSi4C} several values are compiled which compare the quality of
the geometry optimization and relative energies provided by various levels of
theory.  A set of eight bond lengths (four from each Si$_4$C structure) and a
set of ten angles and dihedrals (five from each Si$_4$C structure) are used to
produce MUEs corresponding to the equilibrium bond lengths and angles.  As
before, these MUEs are compiled by comparing against parameters determined at
the highest level of theory employed, which in this case was CCSD(T)/cc-pVTZ.
These quantities show similar trends to those reported in Table \ref{tabsmalls}.
While some DFT methods in conjunction with the cc-pVDZ basis can provide
efficient approaches for obtaining relatively good geometrical parameters, only
the LCCD/cc-pVTZ method is accurate enough to produce MUEs of bond lengths and
angles within 1.0 pm and 1.0 angles of the CCSD(T)/VTZ benchmark.

Also reported in Table \ref{tabSi4C} are the relative energies of each isomer
pair, as computed in two ways.  The first ($\Delta$E(opt)) employed the same
level of theory as is used for the optimization and the second
($\Delta$E(Best)) applied the composite approach given by Eq. \ref{eqcomp} to
the same optimized structures. The $\omega$B97X-D, B3LYP, and LCCD methods alone
failed to unambiguously separate the energies of the two isomers.  The M11 and
MBPT(2) approaches predicted correct qualitative isomer energy orderings, but
excluding CCSD(T)/cc-pVDZ none of the optimization approaches successfully
predicted the energy difference to within 1 kcal/mol of the target
CCSD(T)/cc-pVTZ value. 

\begin{figure}
\includegraphics[width=0.4\paperwidth]{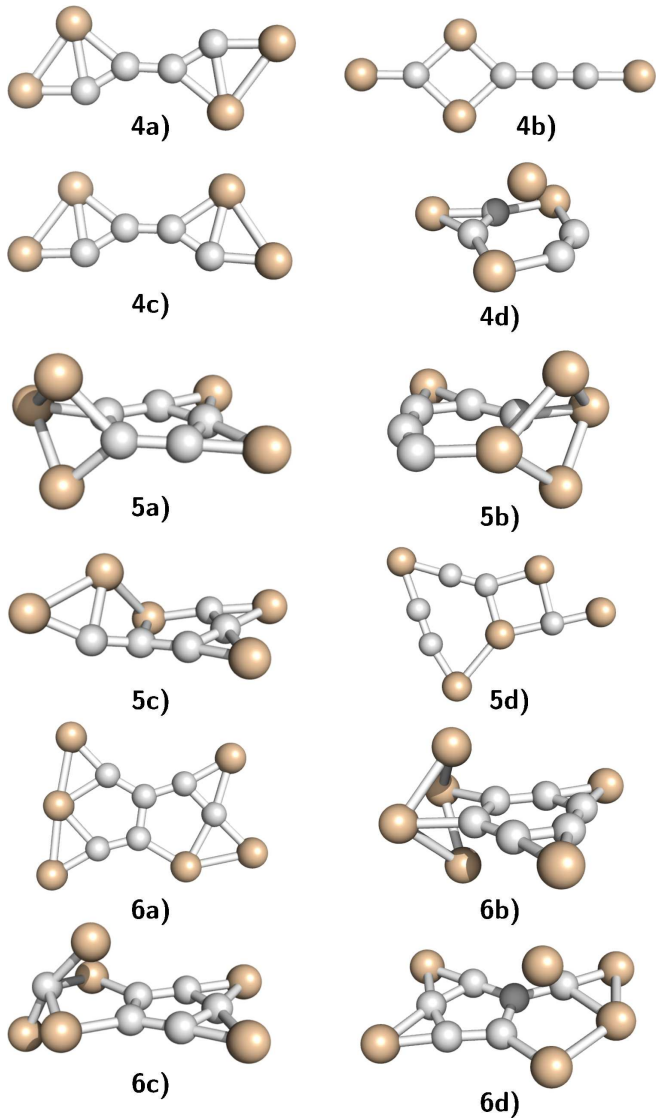}
\caption{\label{456structs}Schematic structures for the four lowest Si$_n$C$_n$
($n=4,\dots,6$) clusters, MBPT(2)/cc-pVTZ optimized coordinates are provided in
the supplemental material.\cite{jcpsupp}}
\end{figure}

Summarizing our findings for the small Si$_n$C$_m$ clusters, reliable geometries
are generally obtained using wave-function methods together with the cc-pVTZ
basis set.  When such methods become computationally intractable, slightly less
robust structures can be obtained in a more efficient manner using DFT methods
with the cc-pVDZ basis set.  While none of the optimization methods tested here
reliably predicted CCSD(T) isomer energy orderings to within chemical accuracy,
the M11 and MBPT(2) approaches were found to at least provide an unambiguous and
correct energy ordering of the Si$_4$C isomers.  Of these two methods only the
MBPT(2) structures result in energy differences within 10\% of benchmark values
when the high-accuracy composite method is subsequently applied.  Thus, while
MBPT(2) may not provide raw Si$_n$C$_m$ energies matching the accuracy of CC
theory, the MBPT(2) method is able to provide both (1) reliable structures for
performing subsequent high-accuracy calculations and (2) raw energies which give
an unambiguous and qualitatively correct isomer energy orderings.  This led us
to choose MBPT(2)/cc-pVTZ as our default method for performing geometry
optimizations on the larger clusters in the following sections. 

\subsection{Accurate Isomers Energy orderings for Si$_n$C$_n$ clusters with {$n$ = 4, 5, 6} }
%\subsection{Accurately Ordering Medium-Sized Clusters by Energy}

In addition to benchmarking smaller silicon carbide systems, we are able to
employ high level coupled-cluster methods on the medium sized Si$_n$C$_n$
clusters.  Previous calculations\cite{DBHm2013} have performed a systematic
search of the configuration space for $n=4,\dots,12$ and identified four lowest
structures for each cluster size to be close in energy using B3LYP.  In order to
validate and benchmark the relative isomer energies the $n=$4, 5 and 6 clusters
are selected for re-examination.  Optimized geometries and hessians are computed
with MBPT(2) (for the coupled-cluster and DH-DFT calculations) or DFT (each
single point is evaluated at the geometry predicted by that DFT functional)
using the cc-pVTZ basis set.  Schematic diagrams of the twelve silicon carbide
clusters can be found in Fig. \ref{456structs}. The MBPT(2)/cc-pVTZ optimized
geometries for all twelve Si$_{n}$C$_{n}$ clusters can be found in the
supplemental material.\cite{jcpsupp}

\begin{figure}[!t]
\includegraphics[width=0.4\paperwidth]{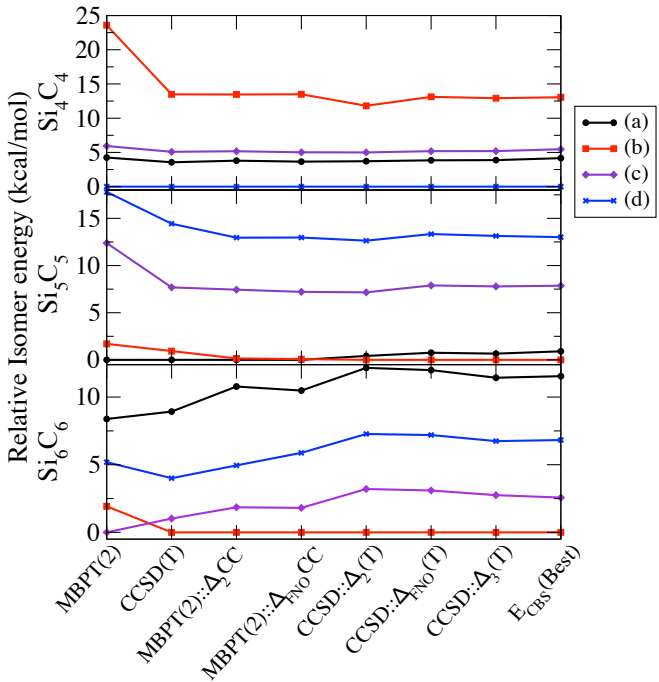}
\caption{\label{456plot}Relative isomer energies for the four lowest ${\rm Si}_n{\rm C}_n$ ($n=4,5,6$) clusters.}
\end{figure}

The $E_{\rm CBS}({\rm Best})$, MBPT(2)::$\Delta_2$CC, and
CCSD::$\Delta_n$(T) relative isomer energies of the ${\rm Si}_n{\rm C}_n$
($n=4,5,6$) clusters are reported in Table \ref{midCBStable} using the MBPT(2)
geometries.  From a pragmatic perspective, the $E_{\rm CBS}({\rm Best})$ (Eq.
\ref{eqcomp}) composite energy does not differ significantly from the
CCSD::$\Delta_n$(T) results that require a computationally expensive cc-pV5Z SCF
calculation.  In Fig. \ref{456plot} the relative isomer energies are plotted to
demonstrate their convergence as the quality of the composite model is
increased.  From Fig. \ref{456plot} it is readily apparent that calculating the
$\Delta$(T) contribution using the cc-pVDZ basis or FNO with the cc-pVTZ basis
is nearly sufficient to obtain quantitative results.  Using only the MBPT(2) CBS
energies with small bases and FNO $\Delta {\rm CC}$ corrections is not
sufficient to completely reproduce the ${\rm Si}_5{\rm C}_5$ predicted isomer
ordering despite having excellent agreement for the other clusters.  As the
difference between 5(a) and 5(b) is $\sim 0.75$ kcal/mol this discrepancy is not
unexpected as it would be overly generous to assign that level of accuracy to an
approximate CBS model applied to a system of this size.

The DH-DFT relative isomer energies compared to the $E_{\rm CBS}({\rm Best})$
reference value are given in Table \ref{midDHtable}.  The DH-DFT single point
energies are obtained at the MBPT(2)/cc-pVTZ geometry and include the
corresponding MBPT(2)/cc-pVTZ ZPE correction.  Overall the DH-DFT results track
the reference values very well, and are a significant improvement over canonical
MBPT(2)/cc-pVQZ (see Table \ref{midDHtable}).  The DSD-PBEP86 flavor performed
notably well across the 12 reference values.

Presented in Table \ref{midDFTtable} are the DFT relative isomer energies as
computed with lower rung functionals.  The results with the various functionals
are mixed, with B3LYP incorrectly predicting the relative energies for any of
the mid sized clusters with an RMS (MAX) error of 7 (12.5) kcal/mol while the
recently developed\cite{jin2016} cam-QTP(0,1) and $\omega$B97X-D functionals
correctly predict the energy orders in some cases with more acceptable RMS (MAX)
errors of 2.8 (7.2) and 2.3 (4.5) kcal/mol respectively.  Specifically the ${\rm
Si}_5{\rm C}_5$ cluster is a difficult case for all the DFT functionals where
the 5(a)-5(b) gap is somewhat greater than the coupled-cluster result and in the
wrong order where 5(a) is predicted to be preferred.  From an
accuracy/efficiency argument the DH-DFT methods provide qualitatively accurate
orderings with reasonably quantitative accuracy when compared with our best CBS
coupled-cluster results. 

\subsection{Isomer Energies of Si$_{12}$C$_{12}$}

\begin{figure}
\includegraphics[width=0.4\paperwidth]{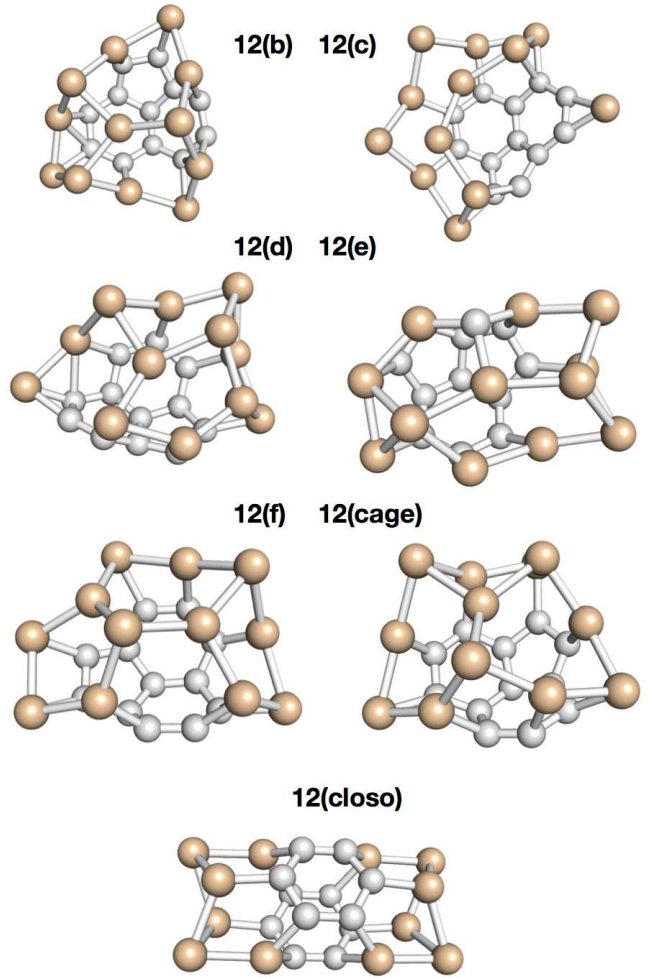}
\caption{\label{12structs}Schematic structures for the seven lowest Si$_{12}$C$_{12}$ clusters, MBPT(2)/cc-pVTZ optimized coordinates are provided in the supplemental material.\cite{jcpsupp}}
\end{figure}

The reliability of various levels of theory for obtaining the optimized
geometries of smaller silicon carbon clusters are assessed in the previous
section, providing a highly accurate framework of model CBS energy decomposition
schemes that will now be used to address the question of isomer ordering and
thermodynamic stability of the Si$_{12}$C$_{12}$ clusters.  We focus our
attention on six cage like configurations\cite{DBHm2013} and the highly
symmetric and compact {\it closo} structure\cite{DBjcp2015} previously reported
to be the lowest preferred isomers of Si$_{12}$C$_{12}$.  Illustrations of these
seven structures are given in Fig. \ref{12structs}.

Optimized geometries and hessians are computed using DFT/cc-pVTZ and
MBPT(2)/cc-pVTZ, the latter used in the calculation of the model CBS energy and
for the DH-DFT calculations.  The MBPT(2)/cc-pVTZ optimized geometries for all
seven Si$_{12}$C$_{12}$ clusters can be found in the supplemental
material.\cite{jcpsupp}  We forgo computing the $E^5_\infty({\rm SCF})$
extrapolated energy as it is clear from Table \ref{midCBStable} that the
$E^4_\infty({\rm SCF})$ is quite sufficient for the larger clusters.  The two
most computationally challenging aspects of the calculation that remain are the
MBPT(2)/cc-pVTZ hessian and CCSD/cc-pVQZ single point energy.  The hessian is
particularly onerous due to the lack of symmetry in most of the cage structures,
resulting in many degrees of freedom.  The CCSD/cc-pVQZ single point energy is a
large calculation with $\sim 1400$ basis functions included.  Fortunately
calculations of this size are becoming
routine\cite{molt2013,byrd2014-a,byrd2015-a,jin2015} for the parallel ACES
programs.

The relative isomer energies are presented in Tables \ref{largeDFTtable},
\ref{largeDHtable}, and \ref{largeCBStable}.  As computed by DFT, the relative
energy between the {\it closo} and various cage structures ranges evenly from
$\sim 4$ to $\sim 20$ kcal/mol with the {\it closo} predicted to be preferred.
The M06-2X and M11 functionals are outliers in this set, with a {\it closo}-cage difference
nearing $\sim 12$ kcal/mol.  A computed difference of $\sim 4$ kcal/mol for a
system this size using DFT is less definitive than otherwise would be desired,
but is none the less suggestive just as previous calculations
concluded.\cite{DBHm2013,DBjcp2015}  This trend is amplified when examining the
DH-DFT results, where the {\it closo}-cage isomer energy gap now $\sim 10$
kcal/mol.  It should be noted that these large Si$_{12}$C$_{12}$ DH-DFT
calculations were performed on commodity workstations

Turning to the proposed CBS models the difference between the  {\it closo} and
other cage structures remains now at $\sim 14$ kcal/mol, a much stronger
indication that the {\it closo} is the Si$_{12}$C$_{12}$ thermodynamically
preferred isomer.  It is evident from Table \ref{largeCBStable} that the simple
CBS model, MBPT(2)::$\Delta_{\rm FNO}$CC, is entirely sufficient for the
selected Si$_{12}$C$_{12}$ structures to decisively predict the large difference
between the {\it closo} and other cages.  Furthermore this approach is
significantly faster to compute and suggests that more approximate approaches
using MBPT(2) as the primary source of correlation could be very successful when
applied to larger silicon carbide systems.  The claim that MBPT(2) correlation
with small corrections from higher-order contributions is sufficient for the
larger SiC clusters is supported by the very good agreement of the DH-DFT
results as compared to the CCSD::$\Delta_3$(T) reference value.  

While CCSD/cc-pVQZ energies are obtainable for the seven structures considered
here, such a calculation might not be practical for much larger systems, much
less their infinite polymer analogues.  Hence it is useful that the $\Delta_{\rm
FNO}$CC correction to the MBPT(2) extrapolated energy is quite sufficient to
obtaining accurate coupled-cluster relative energies as such a calculation is
quick to perform using more readily available hardware (in our case the FNO
calculations were performed on the HiPerGator 2.0 system).  Should even the FNO
be impractical, then qualitative and quantitative energies can nonetheless be
obtained from a DH-DFT calculation. The alternative of using the M11 functional
is also viable, as it is the only tested DFT functional to get close to the coupled-cluster
energy ordering and spacing with the caveat that the M11 functional was unable to 
correctly predict the smaller (${\rm Si}_n{\rm C}_n$, $n\le 6$) siliconcarbon clusters.

\section{Conclusions}    

In this work we have predicted the lowest energy isomer for  some silicon
carbide clusters, ${\rm Si}_{n}{\rm C}_{m}$ ($m,n\le 12$), using extensive
coupled-cluster calculations including converged estimates to the complete basis
set limit.  Our computed isomer order and energy differences for the ${\rm
Si}_{n}{\rm C}_{n}$ ($4\le n\le 6$) clusters (see Table \ref{midCBStable})
differs significantly from previous B3LYP results.\cite{DBHm2013}  The source of
the DFT discrepancy with our best CC based calculations is attributed to the
inadequacies of B3LYP as most of the alternative DFT functions examined here are
far more accurate (Table \ref{midDFTtable}).  In particular the cam-QTP(0,1)
density functional, based on a reparametrization of cam-B3LYP, outperforms the
canonical functional.  Additionally we have included double-hybrid DFT
isomerization energies, which can be considered as a reasonable compromise
between accuracy, theoretical rigor and efficiency.  In addition to computing a
best possible CBS coupled-cluster isomerization energy, we also provide a
systematic study (see Fig. \ref{456plot}) of the convergence for various
approximate composite energy models to clearly illustrate the basis set and
perturbation order important to describing small SiC clusters.  

The {\it closo} ${\rm Si}_{12}{\rm C}_{12}$ structure is confirmed to be the
lowest preferred isomer, consistent with previous B3LYP
calculations.\cite{DBjcp2015}  However most DFT functionals do not accurately reproduce
the CBS coupled-cluster predicted isomer orderings and energy differences for
the various cage structures.  The best possible CBS coupled-cluster composite
model, $E_{\rm CBS}({\rm Best})$ (Eq. \ref{eqcomp}), entails a 4-5$\zeta$ SCF,
and 3-4$\zeta$ CCSD extrapolation with a 3$\zeta$ (T) single point energy 
performed at the MBPT(2)/cc-pVTZ computed geometry.  The $E_{\rm CBS}({\rm
Best})$ calculations include a CCSD/cc-pVQZ calculation (with $\sim 1400$ basis
functions) which we are able to easily perform even for the largest SiC clusters
considered here due to the efficiency of the massively parallel Aces quantum
chemistry packages.  

More pragmatically however, utilizing standard DFT to compute silicon carbon
isomer energies is shown to be highly unreliable, with RMS errors of $\sim 10$
kcal/mol even for the cam-QTP(0,1) and $\omega$B97X-D functionals which were
performed for smaller clusters.  In order to obtain accurate isomer predictions
it is found that a large basis MBPT(2) calculation with some portion of post
MBPT(2) correlation energy is necessary for $\sim 1$ kcal/mol accuracy.  Along
this line, the double-hybrid density functional theory DSD-PBEP86-D3 and
PWPB95-D3 (using the def2-QZVP basis) methods perform well, as do {\it ab
initio} CBS MBPT(2) results corrected by a FNO coupled-cluster calculation.
Both of these more pragmatic composite energies only require modest
computational resources for even the largest silicon carbon clusters considered
here.

Looking forward to much larger ${\rm Si}_n{\rm C}_n$ clusters, alternatives to
performing very large basis CCSD calculations is desirable.  As large basis
MBPT(2) results are becoming increasingly available for very large systems,
through the use of parallel implementations or approximations like the
resolution-of-the-identity, composite CBS energies or alternatives like DH-DFT
are viable when accurate ground state energetics are required.  The relative
errors for the SiC clusters examined here between the approximate models and our
best results are on the order of $1-2$ kcal/mol, leaving the choice in which
calculation to perform is between a balance of theoretical rigor and efficiency.
Our conservative recommendation for future isomerization studies of very large
${\rm Si}_n{\rm C}_m$ clusters is the ${\rm MBPT(2)}::\Delta_{\rm FNO}{\rm CC}$
model (Eq. \ref{fnocbs}), which favors theoretical rigor as increasing system
size can occasionally lead to unanticipated complications with DFT based models.
In cases where DH-DFT or wave-function methods are not affordable, 
it was shown here that M11 more frequently provides an adequate description of the
structure and relative energetics of SiC clusters as compared with other commonly-used
standard DFT functionals.

\section{Acknowledgements}

This work was supported by funding from the U.S. Department of Defense High
Performance Computing Modernization Program and a grant of computer time at the
U.S. NAVY and U.S. Army Research Lab DoD Supercomputing Resource Centers.  
Additionally, JJL was supported in part by an appointment to the Faculty Research Participation Program at U.S. Air Force Institute of Technology (AFIT), administered by the Oak Ridge Institute for Science and Education through an interagency agreement between the U.S. Department of Energy and AFIT.
%Additionally, JJL
%was supported in part by an appointment to the
%Internship/Research Participation Program at the Air Force Institute of
%Technology, administered by the Oak Ridge Institute for Science and Education
%through an interagency agreement between the U.S. Department of Energy and EPA.
The views expressed in this work are those of the authors and do not reflect the
official policy or position of the United States Air Force, Department of
Defense, or the United States Government.

%\bibliography{library}


\begin{table*}[!t]\footnotesize
\caption{\label{tabsmalls}
Mean unsigned errors (MUEs) of bond lengths and bond angles for the ground states of SiC, SiC$_2$, Si$_2$C, and Si$_2$C$_2$, as optimized at various levels of theory.  All values are MUEs reported with respect to the CCSD(T)/aug-cc-pV(Q+d)Z results.}
\begin{tabular}{l c c c c c c c}
\hline
\hline
 & \multicolumn{7}{c}{Basis set\footnote[1]{The basis sets cc-pV$n$Z, aug-cc-pV$n$Z, and aug-cc-pV($n$+d)Z are abbreviated as V$n$Z, AV$n$Z, and AV($n$+d)Z, respectively.}} \\
\cline{2-8}
 Method & VDZ & AVDZ & AV(D+d)Z & VTZ & AVTZ & AV(T+d)Z & AV(Q+d)Z \\
\hline
  \multicolumn{8}{c}{MUE of optimized bond lengths in pm} \\
\hline
$\omega$B97X-D  &0.8    &1.0    &1.1    &1.5    &1.6    &1.7    &1.9 \\
B3LYP		&1.3	&1.1	&1.0	&1.0	&1.0	&1.1	&1.1 \\
M11		&0.7	&1.5	&1.1	&1.9	&1.8	&1.7	&1.2 \\
MBPT(2)		&3.9	&4.2	&3.7	&0.8	&0.9	&0.6	&0.3 \\
%DIPSTEOMMBPT(2)	&1.5	&1.8	&1.4	&0.4	&1.4	&1.7	&1.5 \\
LCCD		&3.4	&3.6	&3.1	&0.7	&0.7	&0.7	&1.1 \\
CCSD		&3.5	&3.7	&3.2	&0.5	&0.4	&0.4	&1.0 \\
CCSD(T)		&4.4	&4.5	&4.1	&0.7	&0.9	&0.6	&0.0 \\
\hline
  \multicolumn{8}{c}{MUE of optimized bond angles in degrees} \\
\hline
$\omega$B97X-D  &1.1    &0.8    &0.9    &1.0    &1.0    &1.0    &0.4 \\
B3LYP		&0.8	&0.7	&0.8	&0.9	&0.9	&0.9	&0.8 \\
M11		&0.6	&0.8	&0.3	&0.8	&1.0	&1.0	&0.9 \\
MBPT(2)		&7.4	&8.4	&7.2	&0.7	&0.5	&0.5	&0.7 \\
LCCD		&7.8	&9.9	&9.1	&0.4	&0.2	&0.3	&0.5 \\
%DIPSTEOMMBPT(2)	&1.4	&4.1	&4.0	&0.7	&1.0	&1.1	&2.5 \\
CCSD		&8.5	&10.2	&9.4	&0.4	&0.5	&0.4	&0.3 \\
CCSD(T)		&11.0	&12.7	&12.0	&0.2	&0.5	&0.2	&0.0 \\
\hline
\hline
\end{tabular}
\end{table*}

\begin{table}\footnotesize
\caption{Basis-set convergence of various leading contributions to the
relative energies of the two lowest-lying structural isomers of SiC$_3$, Si$_3$C$_3$, and Si$_4$C.
For the meaning of the quantities see the text. Energies are in kcal/mol.
\label{tabmeds}}
\begin{tabular}{l c c c c }
\hline
\hline
 Energy       & \multicolumn{4}{c}{Basis set}    \\
\cline{2-5}
contribution  & cc-pVDZ & cc-pVTZ & cc-pVQZ & CBS\footnote[1]{Computed as $E^5_\infty({\rm
SCF})$ and $\Delta^4_\infty {\rm CC}$ from Eqs. \ref{scfcbseqn} and \ref{cccbseqn}, respectively.}  \\
\hline
  \multicolumn{5}{c}{SiC$_3$ isomer energy difference, E(c2)-E(c1)} \\
\hline
$E({\rm SCF})$                     &-10.2  &-11.4 &-11.1
&-11.0 \\
$\Delta {\rm CCSD}$                & 14.3  &17.2  & 18.0 &18.5 \\
$\Delta {\rm (T)}$                 &  1.2  & 1.2  &  1.3 & 1.4 \\
${\rm ZPE}$   & -1.1  &-1.3  &      & \\
$\Delta {\rm CV}$  & -0.1  &-0.4  &      & \\
$\Delta{\rm FS}$    & -0.00 &-0.01 &      & \\
$E_{\rm CBS}({\rm Best})$ & & & &  7.6\footnote[2]{See Eq. \ref{eqcomp}.}   \\
\hline
  \multicolumn{5}{c}{Si$_3$C$_3$ isomer energy difference, E(k2)-E(k1)} \\
\hline
$E({\rm SCF})$                     &-10.2  &-7.2  &-6.9 &-6.5  \\
$\Delta {\rm CCSD}$     & 11.9  &14.8  & 15.7 &16.3 \\
$\Delta {\rm (T)}$      &  1.3  & 1.4  &      & \\
${\rm ZPE}$   & -0.7  & -0.8    &      & \\
$\Delta {\rm CV}$      &  0.5  &      &      & \\
$\Delta{\rm FS}$    & -0.08 &      &      & \\
$E_{\rm CBS}({\rm Best})$ & & & & 10.4\footnotemark[2]    \\
\hline
  \multicolumn{5}{c}{Si$_4$C isomer energy difference, E(m2)-E(m1)} \\
\hline
$E({\rm SCF})$                     &-7.7 &-7.4 &-7.4 &-7.3 \\
$\Delta {\rm CCSD}$     & 7.4 & 7.8 & 8.1 & 8.3 \\
$\Delta {\rm (T)}$      & 2.6 & 2.6   &      & \\
${\rm ZPE}$   &-0.3 & -0.4    &      & \\
$\Delta {\rm CV}$      & 0.1 &      &      & \\
$\Delta{\rm FS}$    & 0.01 &      &      & \\
$E_{\rm CBS}({\rm Best})$ & & & &  3.2\footnotemark[2]   \\
\hline
\hline
\end{tabular}
\\
%\textsuperscript{\emph{a}} { CBS SCF values were computed at cc-pV5Z basis set level (and are within 1 mH of cc-pV6Z values). CBS correlation energies were computed using Eq. \ref{cbs} with cc-pVTZ and cc-pVQZ correlation energies.} \\
%\textsuperscript{\emph{b}} { $E_{\rm CBS}({\rm Best})$ values were computed according to Eq.~\ref{eqcomp}.}
\end{table}

\begin{table}[t]\footnotesize
\caption{Measures of the accuracy of various levels of theory for the geometry optimization 
of Si$_4$C isomers. Stuctural parameters, including bond lengths (in pm) and bond angles and dihedrals
(in degrees), are reported as MUEs with respect to benchmark values optimized at the CCSD(T)/cc-pVTZ level.
Isomer energy differences are reported in kcal/mol. 
\label{tabSi4C}
}
\begin{tabular}{l c c c c c c c c c}
\hline
\hline
Method & Basis set
                  &MUE($R_{\mathrm e})$\footnote[1]{The mean unsigned error of equilibrium bond lengths of the optimized m1 and m2 isomers.} 
                  &MUE($\theta_{\mathrm e})$\footnote[2]{The mean unsigned error of equilibrium bond angles of the optimized m1 and m2 isomers.}
                  &$\Delta$E(opt)\footnote[3]{The isomer energy difference [E(m2)-E(m1)], computed using the specified optimization method.} 
                  &$E_{\rm CBS}({\rm Best})$\footnote[4]{Isomer energy difference [E(m2)-E(m1)], computed using Eq. \ref{eqcomp}.}\\
\hline
$\omega$B97X-D&cc-pVDZ&  7.0 &  2.4 & -0.07 &  7.95   \\
              &cc-pVTZ&  2.3   &  2.7 &  0.01                            & 2.39   \\
B3LYP         &cc-pVDZ&  2.4 &  1.0 & -0.02 &  3.85   \\
              &cc-pVTZ&  1.1   &  0.9 & -0.00  &   3.62 \\
M11           &cc-pVDZ&  0.9 &  1.5 &  1.83 &  4.52   \\
              &cc-pVTZ&  2.1   &  1.2 &  1.71  &   4.07 \\
MBPT(2)       &cc-pVDZ&  2.1 &  1.0 &  4.78 &  3.29  \\
              &cc-pVTZ&  1.4   &  1.2 &  5.79  &   3.18 \\
LCCD          &cc-pVDZ&  2.9 &  0.5 & -0.56 & 3.45 \\
              &cc-pVTZ&  0.8   &  0.6 &  0.31  &   3.27 \\
CCSD(T)       &cc-pVDZ&  3.6 &  0.4 &  2.64 &  3.73 \\
              &cc-pVTZ&  0.0   &  0.0 &  3.55  &   3.52 \\
\hline
\hline
\end{tabular}
\end{table}

\begin{table*}[!b]\footnotesize
\caption{\label{midCBStable}Relative energies (in kcal/mol) for the four lowest ${\rm Si}_n{\rm C}_n$ ($n=4,5,6$) clusters using the MBPT(2)/cc-pVTZ reference geometry.  See text for details of the extrapolation methods used.}
\begin{tabular}{lrrrrrrrr}
\hline
\hline
 & 
 & \multicolumn{1}{c}{MBPT(2)::} 
 & \multicolumn{1}{c}{MBPT(2)::}
 & \multicolumn{1}{c}{CCSD::} 
 & \multicolumn{1}{c}{CCSD::}
 & \multicolumn{1}{c}{CCSD::} 
 & 
 \\
  \multicolumn{1}{c}{${\rm Si}_n{\rm C}_n$}
 & \multicolumn{1}{c}{CCSD(T)\footnote{Single point computed with the cc-pVTZ basis}}
 & \multicolumn{1}{c}{$\Delta_2$CC}
 & \multicolumn{1}{c}{$\Delta_{\rm FNO}$CC}
 & \multicolumn{1}{c}{$\Delta_2$(T)}
 & \multicolumn{1}{c}{$\Delta_{\rm FNO}$(T)}
 & \multicolumn{1}{c}{$\Delta_3$(T)}
 & \multicolumn{1}{c}{$E_{\rm CBS}({\rm Best})$}
 \\
\hline
4(a) & 3.56 & 3.80 & 3.66 & 3.71 & 3.85 & 3.87 & 4.16\\
4(b) & 13.49 & 13.47 & 13.49 & 11.81 & 13.12 & 12.92 & 13.06\\
4(c) & 5.09 & 5.17 & 5.02 & 5.00 & 5.18 & 5.19 & 5.46\\
4(d) & 0.00 & 0.00 & 0.00 & 0.00 & 0.00 & 0.00 & 0.00\\
 & & & & & & & & \\
5(a) & 0.00 & 0.00 & 0.00 & 0.41 & 0.76 & 0.66 & 0.90\\
5(b) & 0.93 & 0.15 & 0.08 & 0.00 & 0.00 & 0.00 & 0.00\\
5(c) & 7.68 & 7.44 & 7.21 & 7.16 & 7.89 & 7.79 & 7.87\\
5(d) & 14.43 & 12.95 & 12.97 & 12.64 & 13.33 & 13.14 & 13.01\\
 & & & & & & & & \\
6(a) & 8.93 & 10.78 & 10.48 & 12.16 & 11.99 & 11.43 & 11.54\\
6(b) & 0.00 & 0.00 & 0.00 & 0.00 & 0.00 & 0.00 & 0.00\\
6(c) & 1.02 & 1.85 & 1.81 & 3.21 & 3.10 & 2.75 & 2.57\\
6(d) & 4.00 & 4.95 & 5.88 & 7.27 & 7.20 & 6.75 & 6.82\\
MAX & 2.82 & 1.88 & 1.06 & 1.25 & 0.53 & 0.29 & \\
RMS & 1.24 & 0.66 & 0.57 & 0.52 & 0.26 & 0.15 & \\
\hline
\hline
\end{tabular}
\end{table*}

\begin{table*}[!t]\footnotesize
\begin{tabular}{lrrrrrrrr}
\hline
\hline
 & \multicolumn{1}{c}{B2PLYP}
& \multicolumn{1}{c}{B2GP-PLYP}
& \multicolumn{1}{c}{DSD-BLYP}
& \multicolumn{1}{c}{DSD-PBEP86}
& \multicolumn{1}{c}{PWPB95}\\
 \multicolumn{1}{c}{${\rm Si}_n{\rm C}_n$}
 & \multicolumn{1}{r}{-D3\textsuperscript{\emph{}}}
& \multicolumn{1}{r}{-D3\textsuperscript{\emph{}}}
& \multicolumn{1}{r}{-D3\textsuperscript{\emph{}}}
& \multicolumn{1}{r}{-D3\textsuperscript{\emph{}}}
& \multicolumn{1}{r}{-D3\textsuperscript{\emph{}}} 
 & \multicolumn{1}{c}{MBPT(2)\footnote[1]{Single point computed with the cc-pVQZ basis.}} 
 & \multicolumn{1}{c}{$E_{\rm CBS}({\rm Best})$\footnote[2]{See Eq. \ref{eqcomp}.}}\\
\hline
4(a) & 0.00 & 1.05 & 1.51 & 2.17 & 3.69 & 4.25 & 4.16\\
4(b) & 3.90 & 7.08 & 10.07 & 12.82 & 14.05 & 23.57 & 13.06\\
4(c) & 1.66 & 2.24 & 2.80 & 3.66 & 5.82 & 5.94 & 5.46\\
4(d) & 1.76 & 0.00 & 0.00 & 0.00 & 0.00 & 0.00 & 0.00\\
 & & & & & & & \\
5(a) & 2.28 & 3.02 & 2.80 & 1.04 & 0.00 & 0.00 & 0.90\\
5(b) & 0.00 & 0.00 & 0.00 & 0.00 & 0.21 & 1.70 & 0.00\\
5(c) & 6.92 & 9.08 & 10.24 & 10.44 & 11.06 & 12.38 & 7.87\\
5(d) & 8.73 & 11.90 & 13.04 & 15.56 & 18.75 & 17.80 & 13.01\\
 & & & & & & & \\
6(a) & 2.58 & 5.70 & 6.14 & 9.38 & 13.41 & 8.38 & 11.54\\
6(b) & 0.38 & 0.00 & 0.00 & 0.00 & 0.00 & 1.92 & 0.00\\
6(c) & 0.00 & 0.39 & 1.03 & 2.75 & 4.17 & 0.00 & 2.57\\
6(d) & 3.45 & 4.40 & 5.14 & 6.02 & 6.65 & 5.19 & 6.82\\
MAX & 9.16 & 5.98 & 5.40 & 2.57 & 5.74 & 10.52 & \\
RMS & 4.46 & 3.00 & 2.36 & 1.47 & 2.07 & 3.88 & \\
\hline
\hline
\end{tabular}
\caption{\label{midDHtable}Relative energies (in kcal/mol) for the four lowest ${\rm Si}_n{\rm C}_n$ ($n=4,5,6$) clusters using density-fitted resolution-of-the-identity double-hybrid DFT as compared to $E_{\rm CBS}({\rm Best})$ at the MBPT(2)/cc-pVTZ reference geometry.}
\end{table*}

\begin{table*}[!t]\footnotesize
\begin{tabular}{lrrrrrrrr}
\hline
\hline
 \multicolumn{1}{c}{${\rm Si}_n{\rm C}_n$}
 & \multicolumn{1}{r}{B3LYP}
& \multicolumn{1}{r}{cam-QTP(0,0)}
& \multicolumn{1}{r}{cam-QTP(0,1)}
%& \multicolumn{1}{r}{BLYP54}
& \multicolumn{1}{r}{$\omega$B97X-D}
& \multicolumn{1}{r}{M06-2X}
& \multicolumn{1}{r}{M11}  
 & \multicolumn{1}{c}{$E_{\rm CBS}({\rm Best})$\footnote{See Eq. \ref{eqcomp}.}}\\
\hline
4(a) & 0.00 & 3.21 & 3.07 & 0.61 & 5.10 & 6.33 & 4.16\\
4(b) & 0.56 & 6.13 & 9.63 & 10.52 & 16.91 & 23.98 & 13.06\\
4(c) & 2.00 & 4.63 & 4.36 & 1.94 & 6.84 & 7.79 & 5.46\\
4(d) & 10.14 & 0.00 & 0.00 & 0.00 & 0.00 & 0.00 & 0.00\\
\\
5(a) & 0.00 & 0.00 & 0.00 & 0.00 & 0.00 & 0.00 & 0.90\\
5(b) & 1.16 & 6.06 & 7.15 & 4.51 & 5.07 & 4.80 & 0.00\\
5(c) & 2.58 & 5.39 & 6.83 & 7.50 & 11.93 & 11.17 & 7.87\\
5(d) & 1.96 & 13.27 & 15.69 & 15.86 & 23.36 & 25.62 & 13.01\\
\\
6(a) & 0.00 & 16.50 & 15.17 & 11.94 & 18.40 & 20.00 & 11.54\\
6(b) & 2.30 & 0.00 & 0.00 & 0.00 & 0.00 & 0.00 & 0.00\\
6(c) & 1.30 & 5.06 & 4.09 & 3.45 & 5.86 & 6.30 & 2.57\\
6(d) & 3.18 & 10.34 & 9.09 & 6.93 & 8.74 & 9.95 & 6.82\\
MAX & 12.49 & 6.93 & 7.15 & 4.51 & 10.35 & 12.61 & \\
RMS & 7.04 & 3.37 & 2.81 & 2.27 & 4.37 & 5.90 & \\
\hline
\hline
\end{tabular}
\caption{\label{midDFTtable}Relative energies (in kcal/mol) for the four lowest ${\rm Si}_n{\rm C}_n$ ($n=4,5,6$) clusters using DFT as compared to $E_{\rm CBS}({\rm Best})$ at the MBPT(2)/cc-pVTZ reference geometry. 
}
\end{table*}

\begin{table*}[!t]\footnotesize
\begin{tabular}{lrrrrrrrr}
\hline
\hline
 \multicolumn{1}{c}{${\rm Si}_n{\rm C}_n$}
 & \multicolumn{1}{r}{B3LYP}
& \multicolumn{1}{r}{cam-QTP(0,0)}
& \multicolumn{1}{r}{cam-QTP(0,1)}
%& \multicolumn{1}{r}{BLYP54}
& \multicolumn{1}{r}{$\omega$B97X-D}
& \multicolumn{1}{r}{M06-2X}
& \multicolumn{1}{r}{M11}  
& \multicolumn{1}{c}{CCSD::$\Delta_3$(T)\footnote[2]{See Eq. \ref{cccbs3}.}}\\
\hline
12(b) & 6.52 & 3.00 & 6.47 & 6.59 & 28.54 & 17.85 & 19.21\\
12(c) & 7.06 & 2.36 & 5.12 & 6.99 & 19.48 & 18.34 & 21.74\\
12(d) & 10.13 & 8.75 & 12.75 & 12.65 & 24.12 & 20.55 & 22.03\\
12(e) & 17.12 & 18.72 & 19.67 & 16.17 & 9.55 & 19.53 & 17.44\\
12(f) & 17.06 & 19.04 & 21.22 & 19.85 & 25.55 & 31.86 & 34.10\\
12(cage) & 3.60 & 0.00 & 3.52 & 4.22 & 24.59 & 12.24 & 13.83\\
12(closo) & 0.00 & 1.60 & 0.00 & 0.00 & 0.00 & 0.00 & 0.00\\
MAX & 17.04 & 19.38 & 16.62 & 14.75 & 10.75 & 3.40 & \\
RMS & 11.42 & 13.29 & 10.70 & 10.43 & 7.05 & 1.98 & \\
\hline
\hline
\end{tabular}
\caption{\label{largeDFTtable}Relative energies (in kcal/mol) for the seven lowest ${\rm Si}_{12}{\rm C}_{12}$ clusters using DFT as compared to CCSD::$\Delta_3$(T) at the MBPT(2)/cc-pVTZ reference geometry.}
\end{table*}

\begin{table*}[!b]\footnotesize
\caption{\label{largeDHtable}Relative energies (in kcal/mol) for the seven
lowest ${\rm Si}_{12}{\rm C}_{12}$ clusters using density-fitted
resolution-of-the-identity double-hybrid DFT as compared to CCSD::$\Delta_3$(T)
at the MBPT(2)/cc-pVTZ reference geometry.}
\begin{tabular}{lrrrrrrr}
\hline
\hline
 & \multicolumn{1}{c}{B2PLYP}
& \multicolumn{1}{c}{B2GP-PLYP}
& \multicolumn{1}{c}{DSD-BLYP}
& \multicolumn{1}{c}{DSD-PBEP86}
& \multicolumn{1}{c}{PWPB95}\\
 \multicolumn{1}{c}{${\rm Si}_n{\rm C}_n$}
 & \multicolumn{1}{r}{-D3}
& \multicolumn{1}{r}{-D3}
& \multicolumn{1}{r}{-D3}
& \multicolumn{1}{r}{-D3}
& \multicolumn{1}{r}{-D3}
 & \multicolumn{1}{c}{MBPT(2)\footnote[1]{Single point computed with the cc-pVQZ basis.}}
 & \multicolumn{1}{c}{CCSD::$\Delta_3$(T)\footnote[2]{See Eq. \ref{cccbs3}.}}\\
\hline
12(b) & 17.24 & 19.33 & 21.81 & 20.57 & 18.86 & 28.54 & 19.21\\
12(c) & 21.32 & 22.45 & 23.22 & 21.75 & 20.61 & 19.48 & 21.74\\
12(d) & 19.47 & 21.48 & 23.10 & 22.32 & 21.92 & 24.12 & 22.03\\
12(e) & 15.61 & 14.81 & 15.03 & 14.97 & 18.24 & 9.55 & 17.44\\
12(f) & 28.91 & 31.00 & 31.71 & 31.50 & 34.54 & 25.55 & 34.10\\
12(cage) & 12.93 & 14.96 & 17.33 & 16.22 & 15.25 & 24.59 & 13.83\\
12(closo) & 0.00 & 0.00 & 0.00 & 0.00 & 0.00 & 0.00 & 0.00\\
MAX & 5.19 & 3.10 & 3.50 & 2.60 & 1.42 & 10.75 & \\
RMS & 2.44 & 1.63 & 2.20 & 1.71 & 0.78 & 7.05 & \\
\hline
\hline
\end{tabular}
\end{table*}

\begin{table*}[!t]\footnotesize
\begin{tabular}{lrrrrrrrr}
\hline
\hline
 & 
 & \multicolumn{1}{c}{MBPT(2)::} 
 & \multicolumn{1}{c}{MBPT(2)::}
 & \multicolumn{1}{c}{CCSD::} 
 & \multicolumn{1}{c}{CCSD::}
 & \multicolumn{1}{c}{CCSD::} 
 & 
 \\
  \multicolumn{1}{c}{${\rm Si}_n{\rm C}_n$}
 & \multicolumn{1}{c}{CCSD(T)\footnote[1]{Single point computed with the cc-pVTZ basis}}
 & \multicolumn{1}{c}{$\Delta_2$CC}
 & \multicolumn{1}{c}{$\Delta_{\rm FNO}$CC}
 & \multicolumn{1}{c}{$\Delta_2$(T)}
 & \multicolumn{1}{c}{$\Delta_{\rm FNO}$(T)}
 & \multicolumn{1}{c}{$\Delta_3$(T)}
\\
\hline
12(b) & 17.85 & 19.08 & 19.41 & 18.85 & 19.49 & 19.21\\
12(c) & 18.34 & 20.43 & 20.55 & 21.47 & 21.74 & 21.74\\
12(d) & 20.55 & 21.69 & 21.84 & 21.79 & 22.07 & 22.03\\
12(e) & 19.53 & 17.75 & 17.73 & 17.80 & 17.37 & 17.44\\
12(f) & 31.86 & 32.79 & 33.32 & 33.91 & 34.13 & 34.10\\
12(cage) & 12.24 & 14.01 & 13.83 & 13.42 & 13.90 & 13.83\\
12(closo) & 0.00 & 0.00 & 0.00 & 0.00 & 0.00 & 0.00\\
MAX & 3.40 & 1.31 & 1.19 & 0.41 & 0.28\\
RMS & 1.98 & 0.73 & 0.56 & 0.29 & 0.11\\
\hline
\hline
\end{tabular}
\caption{\label{largeCBStable}Coupled-cluster CBS relative energies (in
kcal/mol) for the seven lowest ${\rm Si}_{12}{\rm C}_{12}$ clusters using the
MBPT(2)/cc-pVTZ reference geometry.  See text for details of the extrapolation
methods used.}
\end{table*}

\end{document}